\documentclass[twocolumn,prb,citeautoscript,superscriptaddress,8pt]{revtex4-2}

\usepackage{graphicx}
\usepackage{gensymb}
\usepackage{color}
\usepackage[dvipsnames]{xcolor}
\usepackage{float}
\usepackage{upgreek}
\usepackage{bm}
\usepackage{amsmath,amssymb}
\usepackage[colorlinks=true,citecolor=blue,linkcolor=blue]{hyperref}
\setcitestyle{super}

\begin{document}

\title{Persistence and emergence of quantum defects through pressure-induced phase changes}

\author{Alexander J. Healey}
\email{alexander.healey2@rmit.edu.au}
\affiliation{Department of Physics, School of Science, RMIT University, Melbourne, VIC 3001, Australia}

\author{Alan Salek} 
\affiliation{Department of Physics, School of Science, RMIT University, Melbourne, VIC 3001, Australia}

\author{Christopher T.-K. Lew}
\affiliation{Department of Physics, School of Science, RMIT University, Melbourne, VIC 3001, Australia}

\author{Zsolt Benedek}
\affiliation{Department of Physics of Complex Systems, E\"otv\"os Lor\'and University, Budapest, Hungary}

\author{Brett C. Johnson}
\affiliation{Department of Physics, School of Science, RMIT University, Melbourne, VIC 3001, Australia}

\author{Josiah E. Hsi}
\affiliation{Department of Physics, School of Science, RMIT University, Melbourne, VIC 3001, Australia}

\author{Islay O. Robertson}
\affiliation{Department of Physics, School of Science, RMIT University, Melbourne, VIC 3001, Australia}

\author{Kaijian Xing}
\affiliation{Department of Physics, School of Science, RMIT University, Melbourne, VIC 3001, Australia}

\author{Hiroshi Abe}
\affiliation{National Institutes for Quantum Science and Technology (QST), Takasaki, Gunma 370-1292, Japan}

\author{Takeshi Ohshima}
\affiliation{National Institutes for Quantum Science and Technology (QST), Takasaki, Gunma 370-1292, Japan}
\affiliation{Department of Materials Science, Tohoku University, Aoba, Sendai, Miyagi 980-8579, Japan}

\author{Kenji Watanabe}
\affiliation{Research Center for Electronic and Optical Materials, National Institute for Materials Science, Tsukuba, Japan. }

\author{Takashi Taniguchi}
\affiliation{Research Center for Materials Nanoarchitectonics, National Institute for Materials Science, Tsukuba, Japan. }

\author{Mehran Kianinia}
\affiliation{School of Mathematical and Physical Sciences, University of Technology Sydney, Ultimo, New South Wales 2007, Australia}
\affiliation{ARC Centre of Excellence for Transformative Meta-Optical Systems, Faculty of Science, University of Technology Sydney, Ultimo, New South Wales 2007, Australia}

\author{Viktor Iv\'ady}
\affiliation{MTA-ELTE Lend\"ulet ‘Momentum’ NewQubit Research Group, Budapest, Hungary}
\affiliation{Department of Physics of Complex Systems, E\"otv\"os Lor\'and University, Budapest, Hungary}

\author{Igor Aharonovich}
\affiliation{School of Mathematical and Physical Sciences, University of Technology Sydney, Ultimo, New South Wales 2007, Australia}
\affiliation{ARC Centre of Excellence for Transformative Meta-Optical Systems, Faculty of Science, University of Technology Sydney, Ultimo, New South Wales 2007, Australia}

\author{Dougal McCulloch}
\affiliation{Department of Physics, School of Science, RMIT University, Melbourne, VIC 3001, Australia}

\author{Jean-Philippe Tetienne}
\affiliation{Department of Physics, School of Science, RMIT University, Melbourne, VIC 3001, Australia}

\author{David A. Broadway}
\email{david.broadway@rmit.edu.au}
\affiliation{Department of Physics, School of Science, RMIT University, Melbourne, VIC 3001, Australia}

\begin{abstract}
Extreme pressures can transform materials and their properties, but probing these \textit{in-situ} is made challenging by the small sample volumes and access requirements demanded by diamond anvil cells. 
Quantum defects offer a route to local measurements under such conditions, yet their sensing performance can be dictated by pressure-induced changes in their own host material.
On the other hand, pressure may also be harnessed as a tool to engineer and stabilize new quantum defects with emergent functionalities.
Here, we demonstrate both aspects within a unified platform based on optically active spin-pair defects in hexagonal boron nitride (hBN).
As robust quantum sensors under pressure, these spin-1/2 systems retain pressure-independent spin resonances up to 20 GPa while maintaining or even enhancing their optical emission, in stark contrast to the spin-1 boron-vacancy centre in the same material. Simultaneously, we show that compression acts as a means of quantum defect engineering: the starting hBN undergoes an irreversible transformation into wurtzite boron nitride (wBN), during which the defect landscape is reconfigured. Spin-pair sensors are seen to persist across this structural transition, however, depending on the starting material we also observe new, highly fluorescent defects in the wBN phase.
These results establish spin-pair defects in boron nitride as pressure-resilient quantum sensors while highlighting high pressure itself as a versatile pathway for creating and tuning quantum emitters.
\end{abstract}

\maketitle

Extreme pressures are an important tool to engineer the properties of materials, and can lead to emergent behaviors such as high temperature superconductivity and non-trivial magnetic phases~\cite{wang2024c, liu2025c}.
However, options for reliably measuring these phenomena \textit{in-situ} at elevated pressures are limited~\cite{wei2024a}. 
One approach is to use spin-active optical emitters in solid-state materials (quantum defects) that can perform accurate local measurements and have been extensively used to probe condensed matter systems~\cite{casola2018, rovny2025, xue2026}. 
These quantum defects can be placed within the diamond anvils~\cite{hsieh2019, lesik2019,bhattacharyya2024, mandyam2026} or within the sample chamber, directly in contact with the sample~\cite{mu2025,he2025}. The latter approach allows access to near-field signatures, however it could be anticipated that the sensor's host material will also be subject to pressure-induced changes~\cite{gottscholl2021a, mu2025, zhong2025, he2025}. These changes will have consequences for the function of the quantum defects in general, with both the spin and optical responses of these systems potentially tied to properties of the host lattice. Indeed pressure could even drive the evolution, creation, or annihilation of quantum defects. 

\begin{figure*}
    \centering
    \includegraphics[scale=1]{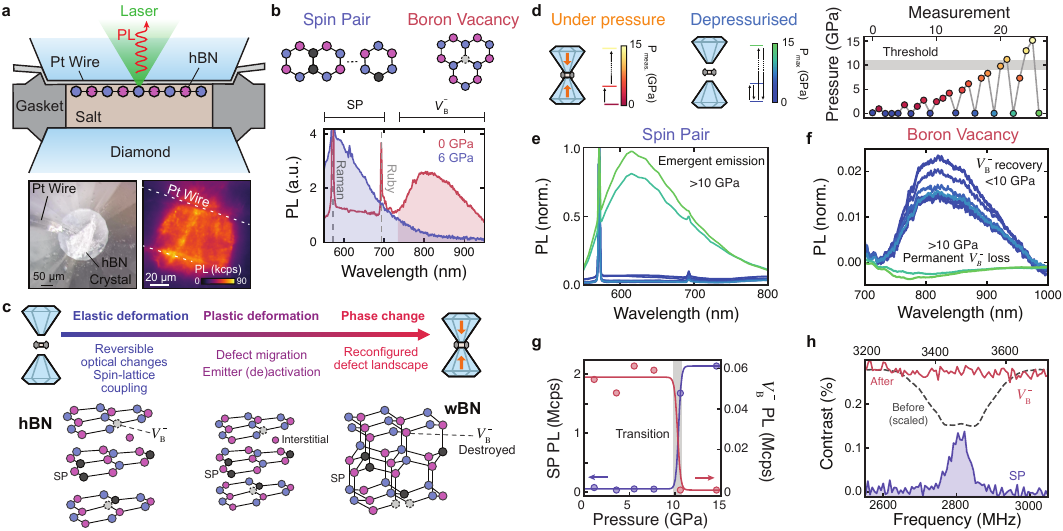}
    \caption{\textbf{Emergent optical behaviour following compression} 
    \textbf{a} Illustration of a quantum-sensing-enabled diamond anvil cell (DAC), with RF control through a Pt wire and optical control and readout performed through one of the anvils. Bottom panels are an optical image and PL image of the same hBN crystal loaded in the DAC.
    \textbf{b} PL spectrum for a typical hBN crystal containing ensembles of both optical-spin defect pairs (SP, broad emission centered at 620 nm) and $V_{\rm B}^-$ defects (emission centered around 800 nm). 
    \textbf{c} Illustration of the deformation process of the hBN crystal with increasing pressure up to a threshold where a structural phase transition occurs to a 3D material.
    \textbf{d} Measurement scheme, where measurements are taken under pressure (orange-yellow data) and after depressurising (blue-green data) including the measurement series used in the following panels. 
    \textbf{e}, \textbf{f} Depressurised optical spectra of the SP and $V_{\rm B}^-$ defects in the same crystal where the background SP signal has been subtracted for clarity in the $V_{\rm B}^-$ case. 
    \textbf{g} Integrated PL for the SP and $V_{\rm B}^-$ as a function of pressure. The phase transition region (grey) is also indicated in \textbf{d}.
    \textbf{h} ODMR spectra recorded at ambient pressure following the $P_{\rm max}=15$ GPa, showing the retention of SP ODMR (blue) and the loss of $V_{\rm B}^-$ ODMR (orange). $V_{\rm B}^-$ data recorded with a 750~nm long pass filter and with zero background magnetic field; SP data collected with a $\sim$200~mT background field and only a laser rejection filter. 
    }
    \label{fig1}
\end{figure*}

Hexagonal boron nitride (hBN) is a prime candidate material to explore these ideas as it is known to host a wide variety of single photon emitters and quantum sensing defects~\cite{gottscholl2020, gottscholl2021, healey2023, vaidya2023,scholten2024, robertson2025}.
These vary in their composition, local defect symmetry, and mechanism for spin state initialisation and readout, meaning an equally diverse range of behaviours versus pressure may be anticipated. 
Furthermore, hBN exhibits a rich phase diagram with its weak interlayer bonding easily modified by pressure, an aspect not yet explored in the context of quantum defects.

In this work we track the evolution of the optical and spin properties of quantum defects in hBN across pressures up to 20~GPa, over which range hBN can undergo a permanent transformation to the wurtzite phase (wBN)~\cite{segura2019,cusco2020,chen2019b}. By considering different hBN starting materials featuring different initial populations of spin defects, we are also able to assess how the defect landscape evolves across this phase transition. This unique platform provides the potential for robust quantum sensing under pressure and introduces a playground for exploring how point defects evolve across structural phase transitions.  \\

\textbf{Probing irreversible changes} \\
To perform quantum measurements at high pressure, we take a commercial diamond anvil cell (DAC), add a Pt wire to deliver radiofrequency (RF) driving for spin state control, and load it with a hBN sample and a NaCl pressure medium (Fig.~\ref{fig1}a). 
The DAC is inserted into a custom-made optical microscopy setup (widefield or confocal depending on the experiment), with laser illumination (generally 532~nm wavelength) delivered and defect photoluminescence (PL) collected through the top diamond anvil (Fig.~\ref{fig1}a). 

We start with hBN crystals containing two varieties of quantum defects: boron vacancies ($V_{\rm B}^-$)~\cite{gottscholl2020,gottscholl2021} produced by electron irradiation and optically active spin-pairs (SPs) related to carbon impurities incorporated during crystal growth~\cite{mendelson2021}. 
The SP and $V_{\rm B}^-$ emission can be separated spectrally (Fig.~\ref{fig1}b), meaning we can directly compare the effect of pressure in the same crystal under identical conditions through optical filtering~\cite{scholten2024}. 
The $V_{\rm B}^-$ emission is known to disappear beyond a few GPa as non-radiative decay pathways become dominant~\cite{mu2025} while the optical emission in the SP can in principle come from arbitrary point defects~\cite{robertson2025} with variable response to pressure, including the possibility for enhancement~\cite{grzeszczyk}. 

Initially, these changes to quantum defect properties are expected to be wholly reversible as the host crystal deforms elastically, however more profound changes are possible at higher pressures when plastic deformation and even the onset of a full phase transition can occur, precipitating defect migration leading to emitter creation or annihilation (Fig.~\ref{fig1}c). 
To probe this transition to permanent structural rearrangement and the consequences for quantum emitters, we recorded the optical response of a crystal containing both SP and $V_{\rm B}^-$ defects following successive pressurisation (pressure $P=P_{\rm max}$) and subsequent depressurisation ($P=0$) cycles, for increasing $P_{\rm max}$ up to 15 GPa (Fig.~\ref{fig1}d). 
Fig.~\ref{fig1}e,f shows an abridged set of measurements (for full series see SI), at ambient pressure ($P_{\rm meas}=0$) following the previous $P_{\rm max}$ (blue to green).
Up to around $P_{\rm max}\approx10$~GPa, changes (SP and $V_{\rm B}^-$ PL increases and decreases respectively, see Fig.~\ref{fig1}b and SI) are predominantly reversible. 
However, past this point, there is an emergent emission from the SP band with an order of magnitude increase in PL (Fig.~\ref{fig1}e), coinciding with the permanent disappearance of $V_{\rm B}^-$ PL (Fig.~\ref{fig1}f).
This can be directly observed in the integrated PL from both emission spectral regions dramatically changing simultaneously (Fig.~\ref{fig1}g).
We conducted \textit{ab-initio} simulations (see SI) to assess the origins of the $V_{\rm B}^-$ disappearance. These indicate that $V_{\rm B}$ defects can irreversibly transform into sp$^3$ point defects near 10~GPa. Substitutional C defects are relatively more robust but can also be modified near this threshold.

This transformation is also observed in the respective optically detected magnetic resonance (ODMR) spectra (which measures the spin frequency of the defect and is discussed later) taken at ambient pressure following $P_{\rm max}=15$ GPa (Fig.~\ref{fig1}h), where the enhanced PL is still spin-active through the SP mechanism while the $V_{\rm B}^-$ spin signature is indeed lost. 
Since SP ODMR contrast is maintained while PL is increased by an order of magnitude, the permanent changes to the hBN crystal translate to meaningfully improved sensor performance in this case.

\begin{figure}
    \centering
    \includegraphics[scale=1]{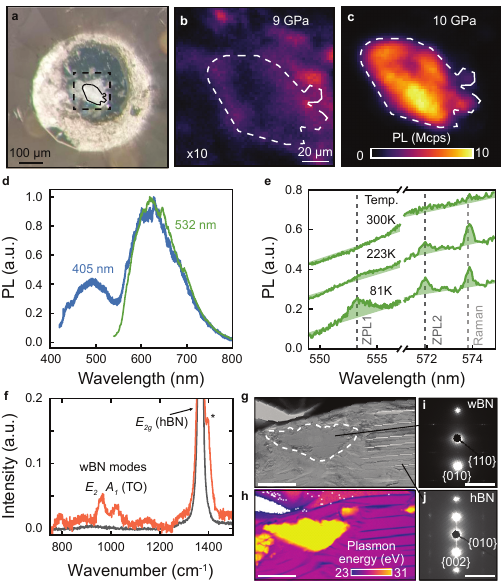}
    \caption{
    \textbf{Emitter creation at phase transition to wBN} 
    \textbf{a} Optical image of hBN crystal following compression. 
    \textbf{b}, \textbf{c} Confocal PL images of hBN crystal at 9~GPa and  10~GPa respectively. 
    \textbf{d} PL spectra obtained with 405~nm (blue) and 532~nm (green) excitation. 
    \textbf{e} Low temperature PL measurements showing the appearance of zero phonon lines at 553 and 572 nm. The main hBN Raman line is indicated at 574 nm, while the wBN Raman signatures (1000~cm$^{-1}$) are around 562 nm (not shown).
    \textbf{f} Raman spectra before (grey) and after (orange) compression showing wBN signatures around 1000~cm$^{-1}$. 
    \textbf{g} Bright field transmission electron microscope image of a cross section of a sample.
    \textbf{h} Electron energy loss spectroscopy image from the region in g, showing the fit plasmon peak position. Scale bars for g,h: 500~nm.
    \textbf{i}, \textbf{j} Selected area electron diffraction patterns taken from wBN ([001] zone axis) and hBN ([100] zone axis) regions respectively. Scale bars: 5~nm$^{-1}$.
    }
    \label{fig:f4}
\end{figure}

\begin{figure}
    \centering
    \includegraphics[width=\linewidth]{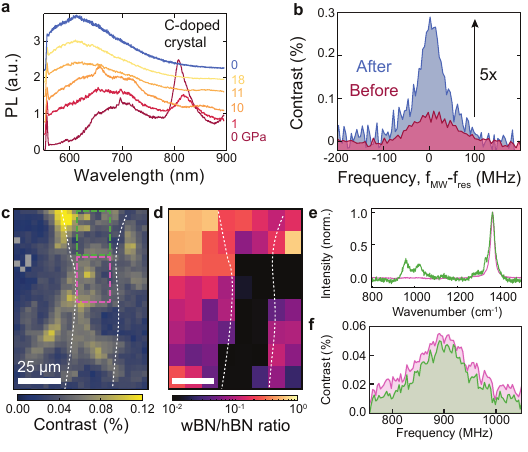}
    \caption{\textbf{Persistence of ODMR through structural phase transition.} 
    \textbf{a} Selected PL spectra obtained from a C-doped hBN crystal at different pressures (full series see SI). 
    \textbf{b} SP ODMR spectra comparing transformed and untransformed hBN material.
    \textbf{c} SP ODMR contrast map of a C-doped hBN crystal recorded at a pressure of 14~GPa. Dotted white lines show location of Pt wire for MW driving.
    \textbf{d} Map of the ratio of wBN and hBN Raman peaks recorded across the region in (c) at ambient pressure upon crystal recovery.
    \textbf{e,f} Raman and ODMR spectra, respectively, averaged over the regions highlighted in (c).}
    \label{fig:f5}
\end{figure}

\textbf{Optical emitter creation during phase transition}\\
The sharp changes in emission properties observed are suggestive of the onset of a structural phase transition. In layered materials like hBN, phase transitions can be strongly dependent on loading conditions or spatially inhomogeneous due to the importance of local shear forces in driving the transition~\cite{ji2012}
In Fig.~\ref{fig:f4}a-c we show an example of this inhomogeneity with a dramatic PL change (two orders of magnitude) confined to one section of the crystal occurring sharply between 9 (Fig.~\ref{fig:f4}b) and 10~GPa (Fig.~\ref{fig:f4}c). 
PL spectra recorded in the bright region reveal a main broad feature around 620~nm when exciting with 532~nm (Fig.~\ref{fig:f4}d). This feature is still evident under 405~nm excitation (along with an additional peak we assign to typical SP emitters~\cite{singh2025a}, unmodified by the transition), suggesting the defects created have a narrow range of zero phonon lines (ZPL).
Indeed low temperature measurements of the bright PL with 532~nm excitation (Fig~\ref{fig:f4}e), shows the emergence of two apparent ZPL at 553~nm and 572~nm. The PL lifetime decay shows clear biexponential behaviour (see SI), further corroborating that the PL enhancement is due to the creation of a confined set of emitters rather than a diverse ensemble.

Raman spectra of the bright region (Fig~\ref{fig:f4}f) detects peaks corresponding to the wurtzite boron nitride (wBN) phase~\cite{cusco2020,segura2019, chen2019b} and a feature at 1395~cm$^{-1}$ (to the right of the normal hBN peak) that may correspond to a compressed hBN phase~\cite{okonai2025}. 
To understand the composition of this mixed phase, we conducted transmission electron microscopy (TEM) on a lamella taken from this region (Fig.~\ref{fig:f4}g).
Analysis of the plasmon peak position via electron energy loss spectroscopy (EELS) reveals extended pockets of wBN, though the majority of the material examined remained in the hBN phase (Fig.~\ref{fig:f4}h). We further establish the crystal phases of these regions through selected area electron diffraction (Fig.~\ref{fig:f4}i,j).
The TEM images show that the unconverted regions exhibit kink band formation~\cite{barsoum2004a}, which is likely a precursor to full conversion to the wurtzite phase. 
Although electron diffraction measurements (Fig.~\ref{fig:f4}j) suggest that the hBN interlayer spacing is preserved on average, a smearing along c$^*$ is present, indicative of disorder along the $c$ axis incurred past the plastic deformation threshold. These structural changes can drive the changes to defect populations and properties observed.
As well as converting to sp$^3$ point defects (locally relieving strain), single vacancies like the $V_{\rm B}^-$ defect may readily annihilate at stacking faults associated with the kink bands~\cite{bai2013} or coalesce to form vacancy clusters or voids in the mixed phase material, all of which are credible candidates for the bright emitters created. 

The new, bright emitters created do not appear to be magnetically active, however ODMR contrast is retained in the the blue-green band (450-550~nm) where this background is avoided (see SI).
The preservation of SP performance at minimum suggests that the component defects are robust to these processes, being likely comprised of single substitutional carbon defects~\cite{robertson2025}.

To concretely test whether SPs can persist through the phase transition to wBN, we consider the compression of a highly carbon-doped bulk hBN sample~\cite{onodera2020}. 
In this case the initial SP population is much higher (related to the carbon impurities~\cite{mendelson2021}) and there are no irradiation-related defects introduced. 
PL measurements (Fig.~\ref{fig:f5}a) show that the emission transitions from a distinct spectral profile~\cite{grzeszczyk} to a featureless (at room temperature) band peaking around 600~nm like other compressed samples (see comparison in SI). 
A sudden change in spectral shape is observed around 10~GPa, coinciding with the inferred hBN-to-wBN transition in the previous crystal.
Raman spectroscopy recorded following decompression (see SI) indeed confirms strong wBN features, indicating a similar phase transition has taken place.  
This time the modulated PL is strongly magnetically active, with SP ODMR (Fig.~\ref{fig:f5}b) enhanced by up to 5 times following the partial transformation to wBN.

Separately, we consider the correlation of an ODMR contrast map recorded \textit{in-situ} at 14~GPa (Fig.~\ref{fig:f5}c) with a Raman map collected at ambient pressure upon sample retrieval (Fig~\ref{fig:f5}d). The Raman map plots the ratio of amplitudes of the peaks corresponding to the wBN $E_2$ and hBN $E_{2g}$ modes and shows regions of varying conversion to the wBN phase. There is not a strong correlation between the ODMR contrast and Raman features beyond the contrast being most consistent near the Pt wire (white dotted lines) where MW driving is strongest. The Pt wire also alters the strain imparted on the hBN crystal and delays the phase transition, though we can identify a region along the wire where there is a strong wBN signal (green box) to compare directly with a region only showing the hBN phase (purple box). Fig.~\ref{fig:f5}e and f show the Raman and ODMR spectra recorded in these regions, highlighting the distinct Raman features and consistency of the ODMR. There are also no strong correlations with the SP PL or other ODMR features (see SI).

These results constitute the strongest evidence that some SP emitters can persist into the wBN phase and highlight the importance of the starting material composition in dictating both the PL emission and sensing performance of the transformed crystals. The recent observation of SPs in cBN~\cite{hsi2026} also corroborates the possibility of some level of agnosticism to the BN phase for some spin-active emitters within the ensembles. 
The permanent changes in PL emission indicate sensitivity to local lattice changes, however this is not strictly a negative change and often results in enhanced PL and ODMR contrast following compression. \\

\begin{figure}
    \centering
    \includegraphics[width=\linewidth]{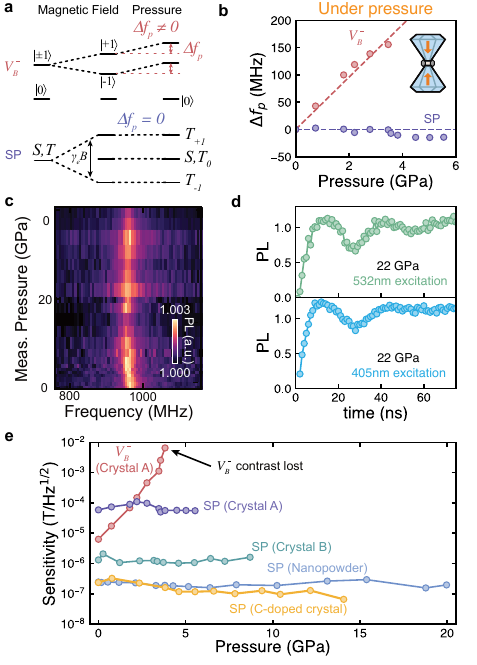}
    \caption{\textbf{Preservation of spin properties up to 20 GPa.}
    \textbf{a} Illustration of the energy levels of the SP system and $V_{\rm B}^-$ (representative of a generic spin-1 system) in response to magnetic field and pressure. 
    \textbf{b} Spin resonance frequency shift versus pressure at a constant magnetic field for SP and $V_{\rm B}^-$.
    \textbf{c} ODMR spectra obtained for a hBN nanopowder sample compressed to 20~GPa and then decompressed back to ambient pressure.
    \textbf{d} Rabi oscillations recorded for a nanopowder sample at a pressure of 22~GPa with 532~nm  (top) and 405~nm (bottom) excitation, with the collection window of  550-650 nm and 450-550 nm respectively.
    \textbf{d} Magnetic sensitivity figure of merit vs pressure for different SP ensembles in nanopowder and bulk crystal samples, again including the $V_{\rm B}^-$ data from Fig.~\ref{fig1} as a comparison. 
    }
    \label{fig:spin}
\end{figure}

\textbf{Quantum sensing under pressure}

The persistence of SP ODMR through dramatic changes in the structure of its host material suggests the potential for maintaining SP sensing function at elevated pressures. Indeed, compared with spin-1 systems such as the $V_{\rm B}^-$ centre, which suffers degraded sensitivity even in the elastic regime~\cite{mu2025,he2025}, the SP system has some features that lessen its intrinsic sensitivity to pressure. 
In the SP mechanism, ODMR occurs as a result of spin-dependent charge transfer between an optical emitter and a separate defect, which can be driven as a metastable weakly coupled spin pair~\cite{robertson2025}. 
Here spin initialisation and readout arise from spin selection rules governing hopping rates rather than local lattice symmetries, and the optical emission can arise from many different point defects spanning the visible band~\cite{singh2025a}, leaving open the possibility for many different behaviours versus pressure. These properties suggest ODMR contrast could be maintained as pressure increases. Further, the effective spin-1/2 sensing state is not directly coupled to the lattice (Fig.~\ref{fig:spin}a)~\cite{robertson2025}, leaving the spin degree of freedom immune to strain gradients that compromise magnetic sensitivity in spin-1 systems~\cite{wang2024c}. 

Starting with a hBN crystal containing both $V_{\rm B}^-$ and SP ensembles, we compare their spin responses under identical conditions. While the $V_{\rm B}^-$ resonance shifts strongly with pressure at a rate of 50 MHz/GPa over the measurable 0-4 GPa range, the SP resonance is unchanged (Fig.~\ref{fig:spin}b), confirming the absence of significant spin-strain coupling. 
Accounting for magnetic field drifts due to screening from the DAC housing (see SI for details), we are able to place an upper bound on the pressure response of the SP ensemble at $<300$~kHz/GPa over the 0-20 GPa range. 
Additionally, the $V_{\rm B}^-$ optical emission disappears beyond a few GPa along with its ODMR contrast~\cite{mu2025}, while the ODMR response of the SP is preserved well above this pressure range, indicating that the state preparation and readout pathways are not significantly modified (see data in SI). 

To benchmark the sensing performance of the SPs under pressure, we consider a nanopowder hBN sample with a high initial population of SP defects. In this case we do not observe a transition to wBN up to the maximum pressure probed of 20~GPa, possibly because the small particle size reduces the prevalence of flake kinking. An irreversible change to the PL spectrum is still evident (see SI), however the spin response (frequency, contrast) is unchanged across a measurement series to a maximum of 20~GPa and gradual decompression back to ambient pressure (Fig.~\ref{fig:spin}c). To assess the coherence of the spins under pressure we measured their response during Rabi driving (Fig~\ref{fig:spin}d) at the maximum pressure (22~GPa). 
The decay of the oscillation is similar to that observed at ambient pressure, indicating that both the intrinsic dephasing and decay rates out of the metastable spin pair state are not strongly modified by pressure~\cite{robertson2025}, and that SPs are therefore amenable to advanced sensing protocol measurements (e.g. dynamical decoupling~\cite{rizzato2023}) under high pressure. 
We repeated the measurement with 405~nm excitation as well as 532~nm and obtained a near-identical response. 
We chose the excitation and collection wavelengths such that two separate ensembles of emitters are measured~\cite{singh2025a}, namely blue-green (450-550 nm) and yellow-red emitters (550-650 nm), and so our measurements here indicate that the pressure-resistant properties are indeed a generic feature of the SP mechanism.

A projection of shot-noise-limited magnetic field sensitivity can be made by measuring the PL emitted, the spin readout contrast, and resonance linewidth. The sensitivity measured versus pressure is plotted in Fig.~\ref{fig:spin}e for various hBN samples. In all cases we find that the SP sensitivity is maintained (and sometimes improved) across the pressure range probed, with the value scaled by the initial properties of the SP ensemble as well as the measurement conditions. Again we include an explicit comparison between SP and $V_{\rm B}^-$ spins (Crystal A; red and purple points). This crystal was initially optimised for $V_{\rm B}^-$ sensitivity however this rapidly degrades as pressure increases, highlighting the fundamental differences between the two sensor classes. We also include the C-doped hBN sample from Fig.~\ref{fig:f5}c, which exhibits a monotonic improvement in sensitivity with pressure as ODMR contrast improves.

The values obtained with SP ensembles are already competitive even with state-of-the-art, optimised NV-DAC implementations~\cite{wang2024c}, with the added advantage of the potential avoidance of sensor-target standoff and freedom to apply strong bias fields along any direction~\cite{scholten2024, healey2023}. While we explicitly highlight hBN samples that do not undergo phase transitions over the pressure range probed (mapping to the ideal, hydrostatic case), the crystal-dependent potential for SP property enhancement through these structural changes shown previously also presents an opportunity for sensing function to be retained in that case. Here we can envision the PL changes established acting as a sentinel for local changes to the host crystal while the spin degree of freedom remains unperturbed, retaining its full magnetic sensing function.\\

\textbf{Conclusion} \\ 
Our results have showcased two complementary aspects revealed by tracking the evolution of quantum defects in hBN with pressure. 
The SP system offers unique opportunities for \textit{in-situ} sensing at high pressures, combining an optical reporter of local defect structure with a magnetic sensor decoupled from lattice properties. These findings establish hBN SP sensors as versatile magnetic probes at high pressure, with the natural integrability of hBN into van der Waals heterostructures offering particular promise for measuring hitherto invisible near-field signatures.

On the other hand, the hBN starting material is permanently modified by pressures on the 10~GPa scale. We have seen that the transition to the wBN phase is marked by dramatic changes in optical properties depending on the starting material. Pressure is therefore a tool for engineering quantum defects that has not been well explored in the literature to date, and our observation of highly fluorescent colour centres in wBN introduces it as a candidate material for quantum technologies going forward. Equally, these optical signatures offer a sharp and highly local probe of the structural phase transition. Given the demonstrated robustness of the SP system to these changes, an intriguing direction is to utilise the magnetic degree of freedom through the hyperfine interaction on the single defect level~\cite{gao2025b} to give a multi-faceted view of the local phase transition. Finally, given the ubiquity of the SP model (already known to describe defects in a range of wide bandgap semiconductors~\cite{robertson2025,vaidya2025}), our work offers a blueprint for studies into other materials, giving a window into the defect-mediated kinetics of their phase transitions. 


\bibliographystyle{MSP}
\bibliography{library}

@Article{liu2025c,
  author    = {Liu, Qiye and Su, Wenjie and Gu, Yue and Zhang, Xi and Xia, Xiuquan and Wang, Le and Xiao, Ke and Zhang, Naipeng and Cui, Xiaodong and Huang, Mingyuan and Wei, Chengrong and Zou, Xiaolong and Xi, Bin and Mei, Jia-Wei and Dai, Jun-Feng},
  title     = {Surprising Pressure-Induced Magnetic Transformations from Helimagnetic Order to Antiferromagnetic State in {{NiI2}}},
  doi       = {10.1038/s41467-025-59561-0},
  issn      = {2041-1723},
  number    = {1},
  pages     = {4221},
  urldate   = {2026-01-13},
  volume    = {16},
  copyright = {2025 The Author(s)},
  journal   = {Nature Communications},
  langid    = {english},
  month     = may,
  publisher = {Nature Publishing Group},
  year      = {2025},
}

@Article{rizzato2023,
  author    = {Rizzato, Roberto and Schalk, Martin and Mohr, Stephan and Hermann, Jens C. and Leibold, Joachim P. and Bruckmaier, Fleming and Salvitti, Giovanna and Qian, Chenjiang and Ji, Peirui and Astakhov, Georgy V. and Kentsch, Ulrich and Helm, Manfred and Stier, Andreas V. and Finley, Jonathan J. and Bucher, Dominik B.},
  title     = {Extending the Coherence of Spin Defects in {{hBN}} Enables Advanced Qubit Control and Quantum Sensing},
  doi       = {10.1038/s41467-023-40473-w},
  issn      = {2041-1723},
  pages     = {5089},
  urldate   = {2025-10-14},
  volume    = {14},
  copyright = {2023 The Author(s)},
  journal   = {Nature Communications},
  langid    = {english},
  month     = aug,
  publisher = {Nature Publishing Group},
  year      = {2023},
}

@Article{gottscholl2021a,
  author    = {Gottscholl, Andreas and Diez, Matthias and Soltamov, Victor and Kasper, Christian and Krau{\ss}e, Dominik and Sperlich, Andreas and Kianinia, Mehran and Bradac, Carlo and Aharonovich, Igor and Dyakonov, Vladimir},
  title     = {Spin Defects in {{hBN}} as Promising Temperature, Pressure and Magnetic Fi Eld Quantum Sensors},
  doi       = {10.1038/s41467-021-24725-1},
  issn      = {2041-1723},
  number    = {4480},
  volume    = {12},
  journal   = {Nat. Commun.},
  publisher = {Springer US},
  year      = {2021},
}

@Article{okonai2025,
  author   = {Okonai, Takara and {Sol{\'i}s-Fern{\'a}ndez}, Pablo and Fukamachi, Satoru and Sun, Haiming and Lee, Yeri and Lin, Yung-Chang and Kato, Toshiaki and Ryu, Sunmin and Suenaga, Kazu and Ago, Hiroki},
  title    = {Anomalous {{Raman}} Signals in Multilayer Hexagonal Boron Nitride Grown by Chemical Vapour Deposition on Metal Foil Catalysts},
  doi      = {10.1039/D5NA00283D},
  issn     = {2516-0230},
  pages    = {10.1039.D5NA00283D},
  urldate  = {2025-11-12},
  journal  = {Nanoscale Advances},
  langid   = {english},
  year     = {2025},
}

@Article{grzeszczyk,
  author    = {Grzeszczyk, Magdalena and Badrtdinov, Danis I. and Watanabe, Kenji and Taniguchi, Takashi and Dreyer, Cyrus E. and R{\"o}sner, Malte and Koperski, Maciej},
  title     = {Orbital {{Geometry-Governed Response}} of {{Pressure-Tunable Quantum Defects}} in {{hBN}}},
  doi       = {10.1002/adfm.75787},
  issn      = {1616-3028},
  pages     = {e75787},
  urldate   = {2026-05-19},
  copyright = {\copyright{} 2026 The Author(s). Advanced Functional Materials published by Wiley-VCH GmbH},
  journal   = {Advanced Functional Materials},
  langid    = {english},
  year      = {2026},
}

@Article{hsieh2019,
  author   = {Hsieh, S. and Bhattacharyya, P. and Zu, C. and Mittiga, T. and Smart, T. J. and Machado, F. and Kobrin, B. and H{\"o}hn, T. O. and Rui, N. Z. and Kamrani, M. and Chatterjee, S. and Choi, S. and Zaletel, M. and Struzhkin, V. V. and Moore, J. E. and Levitas, V. I. and Jeanloz, R. and Yao, N. Y.},
  title    = {Imaging Stress and Magnetism at High Pressures Using a Nanoscale Quantum Sensor},
  doi      = {10.1126/science.aaw4352},
  issn     = {0036-8075, 1095-9203},
  number   = {6471},
  pages    = {1349--1354},
  urldate  = {2026-05-21},
  volume   = {366},
  journal  = {Science},
  langid   = {english},
  month    = dec,
  year     = {2019},
}

@Article{segura2019,
  author     = {Segura, A. and Cusc{\'o}, R. and Taniguchi, T. and Watanabe, K. and Cassabois, G. and Gil, B. and Art{\'u}s, L.},
  title      = {Nonreversible {{Transition}} from the {{Hexagonal}} to {{Wurtzite Phase}} of {{Boron Nitride}} under {{High Pressure}}: {{Optical Properties}} of the {{Wurtzite Phase}}},
  doi        = {10.1021/acs.jpcc.9b06163},
  issn       = {1932-7447, 1932-7455},
  number     = {33},
  pages      = {20167--20173},
  urldate    = {2025-08-27},
  volume     = {123},
  copyright  = {https://doi.org/10.15223/policy-029},
  journal    = {The Journal of Physical Chemistry C},
  langid     = {english},
  month      = aug,
  shorttitle = {Nonreversible {{Transition}} from the {{Hexagonal}} to {{Wurtzite Phase}} of {{Boron Nitride}} under {{High Pressure}}},
  year       = {2019},
}

@Article{rovny2025,
  author    = {Rovny, Jared and Kolkowitz, Shimon and {de Leon}, Nathalie P.},
  title     = {Multi-Qubit Nanoscale Sensing with Entanglement as a Resource},
  doi       = {10.1038/s41586-025-09760-y},
  issn      = {1476-4687},
  number    = {8091},
  pages     = {876--882},
  urldate   = {2025-11-30},
  volume    = {647},
  copyright = {2025 The Author(s), under exclusive licence to Springer Nature Limited},
  journal   = {Nature},
  langid    = {english},
  month     = nov,
  publisher = {Nature Publishing Group},
  year      = {2025},
}

@Article{bhattacharyya2024,
  author  = {Bhattacharyya, P. and Chen, W and Huang, X and Chatterjee, S and Huang, B and Kobrin, B and Lyu, Y and Cui, T and Galli, G and Halperin, B I and Laumann, C R and Yao, N Y},
  title   = {Imaging the {{Meissner}} Effect in Hydride Superconductors Using Quantum Sensors},
  doi     = {10.1038/s41586-024-07026-7},
  pages   = {73-79},
  volume  = {627},
  journal = {Nature},
  year    = {2024},
}

@Article{zhong2025,
  author     = {Zhong, Cheng and Mai, Di and Wang, Yupeng and Wang, He and Dai, Rucheng and Wang, Zhongping and Sun, Xiaoyu and Zhang, Zengming},
  title      = {Ultrasensitive {{Pressure Sensing}} with {{Boron Vacancy Defects}} in {{Hexagonal Boron Nitride}}: {{In-Situ Pressure Imaging}} of {{Two-Dimensional Heterostructures}} under {{High Pressure}}},
  doi        = {10.1021/acsphotonics.5c00670},
  urldate    = {2025-06-16},
  journal    = {ACS Photonics},
  month      = jun,
  publisher  = {American Chemical Society},
  shorttitle = {Ultrasensitive {{Pressure Sensing}} with {{Boron Vacancy Defects}} in {{Hexagonal Boron Nitride}}},
  year       = {2025},
}

@Article{scholten2024,
  author    = {Scholten, Sam C. and Singh, Priya and Healey, Alexander J. and Robertson, Islay O. and Haim, Galya and Tan, Cheng and Broadway, David A. and Wang, Lan and Abe, Hiroshi and Ohshima, Takeshi and Kianinia, Mehran and Reineck, Philipp and Aharonovich, Igor and Tetienne, Jean-Philippe},
  title     = {Multi-Species Optically Addressable Spin Defects in a van Der {{Waals}} Material},
  doi       = {10.1038/s41467-024-51129-8},
  issn      = {2041-1723},
  pages     = {6727},
  urldate   = {2025-06-30},
  volume    = {15},
  copyright = {2024 The Author(s)},
  journal   = {Nature Communications},
  langid    = {english},
  month     = aug,
  publisher = {Nature Publishing Group},
  year      = {2024},
}

@Article{cusco2020,
  author  = {Cusc{\'o}, Ramon and {Pellicer-Porres}, Julio and Edgar, James H. and Li, Jiahan and Segura, Alfredo and Art{\'u}s, Luis},
  title   = {Pressure Dependence of the Interlayer and Intralayer {{E}} 2 g {{Raman-active}} Modes of Hexagonal {{BN}} up to the Wurtzite Phase Transition},
  doi     = {10.1103/PhysRevB.102.075206},
  issn    = {2469-9950, 2469-9969},
  number  = {7},
  pages   = {075206},
  urldate = {2025-08-27},
  volume  = {102},
  journal = {Physical Review B},
  langid  = {english},
  month   = aug,
  year    = {2020},
}

@Article{gao2025b,
  author    = {Gao, Xingyu and Vaidya, Sumukh and Li, Kejun and Ge, Zhun and Dikshit, Saakshi and Zhang, Shimin and Ju, Peng and Shen, Kunhong and Jin, Yuanbin and Ping, Yuan and Li, Tongcang},
  title     = {Single Nuclear Spin Detection and Control in a van Der {{Waals}} Material},
  doi       = {10.1038/s41586-025-09258-7},
  issn      = {1476-4687},
  pages     = {943-949},
  urldate   = {2025-07-10},
  volume    = {643},
  copyright = {2025 The Author(s)},
  journal   = {Nature},
  langid    = {english},
  month     = jul,
  publisher = {Nature Publishing Group},
  year      = {2025},
}

@Article{casola2018,
  author  = {Casola, Francesco and {van der Sar}, Toeno and Yacoby, Amir},
  title   = {Probing Condensed Matter Physics with Magnetometry Based on Nitrogen-Vacancy Centres in Diamond},
  doi     = {10.1038/natrevmats.2017.88},
  number  = {17088},
  volume  = {3},
  journal = {Nat. Rev. Mater.},
  year    = {2018},
}

@Article{mandyam2026,
  author    = {Mandyam, S. V. and Wang, E. and Wang, Z. and Chen, B. and Jayarama, N. C. and Gupta, A. and Riesel, E. A. and Levitas, V. I. and Laumann, C. R. and Yao, N. Y.},
  title     = {Uncovering Origins of Heterogeneous Superconductivity in {{La3Ni2O7}}},
  doi       = {10.1038/s41586-025-10095-x},
  issn      = {1476-4687},
  pages     = {54-60},
  urldate   = {2026-02-25},
  volume    = {651},
  copyright = {2026 The Author(s)},
  journal   = {Nature},
  langid    = {english},
  month     = feb,
  publisher = {Nature Publishing Group},
  year      = {2026},
}

@Article{ji2012,
  author   = {Ji, Cheng and Levitas, Valery I. and Zhu, Hongyang and Chaudhuri, Jharna and Marathe, Archis and Ma, Yanzhang},
  title    = {Shear-Induced Phase Transition of Nanocrystalline Hexagonal Boron Nitride to Wurtzitic Structure at Room Temperature and Lower Pressure},
  doi      = {10.1073/pnas.1214976109},
  issn     = {0027-8424, 1091-6490},
  number   = {47},
  pages    = {19108--19112},
  urldate  = {2026-08-06},
  volume   = {109},
  journal  = {Proceedings of the National Academy of Sciences},
  langid   = {english},
  month    = nov,
  year     = {2012},
}

@Article{he2025,
  author    = {He, Guanghui and Gong, Ruotian and Wang, Zhipan and Liu, Zhongyuan and Hong, Jeonghoon and Zhang, Tongxie and Riofrio, Ariana L. and Rehfuss, Zackary and Chen, Mingfeng and Yao, Changyu and Poirier, Thomas and Ye, Bingtian and Wang, Xi and Ran, Sheng and Edgar, James H. and Zhang, Shixiong and Yao, Norman Y. and Zu, Chong},
  title     = {Probing Stress and Magnetism at High Pressures with Two-Dimensional Quantum Sensors},
  doi       = {10.1038/s41467-025-63535-7},
  issn      = {2041-1723},
  number    = {1},
  pages     = {8162},
  urldate   = {2025-09-21},
  volume    = {16},
  copyright = {2025 The Author(s)},
  journal   = {Nature Communications},
  langid    = {english},
  month     = sep,
  publisher = {Nature Publishing Group},
  year      = {2025},
}

@Article{mu2025,
  author    = {Mu, Z. and Frauni{\'e}, J. and Durand, A. and Cl{\'e}ment, S. and Finco, A. and Rouquette, J. and {Hadj-Azzem}, A. and Rougemaille, N. and Coraux, J. and Li, J. and Poirier, T. and Edgar, J. H. and Gerber, I. C. and Marie, X. and Gil, B. and Cassabois, G. and Robert, C. and Jacques, V.},
  title     = {Magnetic Imaging under High Pressure with a Spin-Based Quantum Sensor Integrated in a van Der {{Waals}} Heterostructure},
  doi       = {10.1038/s41467-025-63580-2},
  issn      = {2041-1723},
  pages     = {8574},
  urldate   = {2025-09-29},
  volume    = {16},
  copyright = {2025 The Author(s)},
  journal   = {Nature Communications},
  langid    = {english},
  month     = sep,
  publisher = {Nature Publishing Group},
  year      = {2025},
}

@Article{wei2024a,
  author     = {Wei, Bin and Lin, Lin and Zhang, Jian and Zhan, Zhiwen and Cheng, Ziwei and Jiang, Junru},
  title      = {In {{Situ Measurement Techniques Using Diamond Anvil Cell}} at {{High Pressure}}--{{Temperature Conditions}}: {{A Review}}},
  doi        = {10.1002/pssr.202300469},
  issn       = {1862-6270},
  number     = {8},
  pages      = {2300469},
  volume     = {18},
  journal    = {physica status solidi (RRL) -- Rapid Research Letters},
  shorttitle = {In {{Situ Measurement Techniques Using Diamond Anvil Cell}} at {{High Pressure}}--{{Temperature Conditions}}},
  year       = {2024},
}

@Article{hsi2026,
  author    = {Hsi, Josiah E. and Robertson, Islay O. and Biswas, Abhijit and Murukeshan, Jishnu and Khabashesku, Valery and Healey, Alexander J. and Grant, Erin S. and Broadway, David A. and Kianinia, Mehran and Aharonovich, Igor and Ajayan, Pulickel M. and Tetienne, Jean-Philippe},
  title     = {Optical {{Spin Defect Pairs}} in {{Cubic Boron Nitride}}},
  doi       = {10.1021/acsphotonics.6c01030},
  issn      = {2330-4022, 2330-4022},
  pages     = {acsphotonics.6c01030},
  urldate   = {2026-07-28},
  copyright = {https://doi.org/10.15223/policy-029},
  journal   = {ACS Photonics},
  langid    = {english},
  month     = jul,
  year      = {2026},
}

@Article{singh2025a,
  author   = {Singh, Priya and Robertson, Islay O. and Scholten, Sam C. and Healey, Alexander J. and Abe, Hiroshi and Ohshima, Takeshi and Tan, Hark Hoe and Kianinia, Mehran and Aharonovich, Igor and Broadway, David A. and Reineck, Philipp and Tetienne, Jean-Philippe},
  title    = {Violet to {{Near-Infrared Optical Addressing}} of {{Spin Pairs}} in {{Hexagonal Boron Nitride}}},
  doi      = {10.1002/adma.202414846},
  issn     = {1521-4095},
  pages    = {2414846},
  urldate  = {2025-02-19},
  volume   = {37},
  journal  = {Advanced Materials},
  langid   = {english},
  year     = {2025},
}

@Article{lesik2019,
  author   = {Lesik, Margarita and Plisson, Thomas and Toraille, Lo{\"i}c and Renaud, Justine and Occelli, Florent and Schmidt, Martin and Salord, Olivier and Delobbe, Anne and Debuisschert, Thierry and Rondin, Lo{\"i}c and Loubeyre, Paul and Roch, Jean-Fran{\c c}ois},
  title    = {Magnetic Measurements on Micrometer-Sized Samples under High Pressure Using Designed {{NV}} Centers},
  doi      = {10.1126/science.aaw4329},
  issn     = {0036-8075, 1095-9203},
  number   = {6471},
  pages    = {1359--1362},
  urldate  = {2026-05-21},
  volume   = {366},
  journal  = {Science},
  langid   = {english},
  month    = dec,
  year     = {2019},
}

@Article{wang2024c,
  author    = {Wang, Mengqi and Wang, Yu and Liu, Zhixian and Xu, Ganyu and Yang, Bo and Yu, Pei and Sun, Haoyu and Ye, Xiangyu and Zhou, Jingwei and Goncharov, Alexander F. and Wang, Ya and Du, Jiangfeng},
  title     = {Imaging Magnetic Transition of Magnetite to Megabar Pressures Using Quantum Sensors in Diamond Anvil Cell},
  doi       = {10.1038/s41467-024-52272-y},
  issn      = {2041-1723},
  pages     = {8843},
  urldate   = {2025-10-28},
  volume    = {15},
  copyright = {2024 The Author(s)},
  journal   = {Nature Communications},
  langid    = {english},
  month     = oct,
  publisher = {Nature Publishing Group},
  year      = {2024},
}

@Article{bai2013,
  author     = {Bai, Xian-Ming and Uberuaga, Blas P.},
  title      = {The {{Influence}} of {{Grain Boundaries}} on {{Radiation-Induced Point Defect Production}} in {{Materials}}: {{A Review}} of {{Atomistic Studies}}},
  doi        = {10.1007/s11837-012-0544-5},
  issn       = {1047-4838, 1543-1851},
  number     = {3},
  pages      = {360--373},
  urldate    = {2026-05-06},
  volume     = {65},
  copyright  = {http://www.springer.com/tdm},
  journal    = {JOM},
  langid     = {english},
  month      = mar,
  shorttitle = {The {{Influence}} of {{Grain Boundaries}} on {{Radiation-Induced Point Defect Production}} in {{Materials}}},
  year       = {2013},
}

@Article{robertson2025,
  author   = {Robertson, Islay O. and Whitefield, Benjamin and Scholten, Sam C. and Singh, Priya and Healey, Alexander J. and Reineck, Philipp and Kianinia, Mehran and Barcza, Gergely and Iv{\'a}dy, Viktor and Broadway, David A. and Aharonovich, Igor and Tetienne, Jean-Philippe},
  title    = {A Charge Transfer Mechanism for Optically Addressable Solid-State Spin Pairs},
  doi      = {10.1038/s41567-025-03091-5},
  issn     = {1745-2473, 1745-2481},
  number   = {12},
  pages    = {1981--1987},
  urldate  = {2025-12-15},
  volume   = {21},
  journal  = {Nature Physics},
  langid   = {english},
  month    = dec,
  year     = {2025},
}

@Article{xue2026,
  author    = {Xue, Ruolan and Maksimovic, Nikola and Dolgirev, Pavel E. and Xia, Li-Qiao and M{\"u}ller, Aaron and Kitagawa, Ryota and Machado, Francisco and Klein, Dahlia R. and MacNeill, David and Watanabe, Kenji and Taniguchi, Takashi and {Jarillo-Herrero}, Pablo and Lukin, Mikhail D. and Demler, Eugene and Yacoby, Amir},
  title     = {Magnon Hydrodynamics in an Atomically Thin Ferromagnet},
  doi       = {10.1126/science.adp2397},
  number    = {6800},
  pages     = {873--878},
  urldate   = {2026-05-22},
  volume    = {392},
  journal   = {Science},
  month     = may,
  publisher = {American Association for the Advancement of Science},
  year      = {2026},
}

@Article{vaidya2025,
  author    = {Vaidya, Sumukh and Gao, Xingyu and Dikshit, Saakshi and Fang, Zhenyao and Llacsahuanga Allcca, Andres E. and Chen, Yong P. and Yan, Qimin and Li, Tongcang},
  title     = {Coherent {{Spins}} in van Der {{Waals Semiconductor GeS}}{\textsubscript{2}} at {{Ambient Conditions}}},
  doi       = {10.1021/acs.nanolett.5c03567},
  issn      = {1530-6984, 1530-6992},
  number    = {39},
  pages     = {14356–14362},
  urldate   = {2025-09-18},
  volume    = {25},
  copyright = {https://creativecommons.org/licenses/by-nc-nd/4.0/},
  journal   = {Nano Letters},
  langid    = {english},
  month     = sep,
  year      = {2025},
}

@Article{mendelson2021,
  author    = {Mendelson, Noah and Chugh, Dipankar and Reimers, Jeffrey R and Cheng, Tin S and Gottscholl, Andreas and Long, Hu and Mellor, Christopher J and Zettl, Alex and Dyakonov, Vladimir and Beton, Peter H and Novikov, Sergei V and Jagadish, Chennupati and Tan, Hark Hoe and Ford, Michael J and Toth, Milos and Bradac, Carlo and Aharonovich, Igor},
  title     = {Identifying Carbon as the Source of Visible Single-Photon Emission from Hexagonal Boron Nitride},
  doi       = {10.1038/s41563-020-00850-y},
  issn      = {1476-4660},
  pages     = {321--328},
  volume    = {20},
  journal   = {Nat. Mater.},
  publisher = {Springer US},
  year      = {2021},
}

@Article{vaidya2023,
  author    = {Vaidya, Sumukh and Gao, Xingyu and Dikshit, Saakshi and Aharonovich, Igor and Li, Tongcang},
  title     = {Quantum Sensing and Imaging with Spin Defects in Hexagonal Boron Nitride},
  doi       = {10.1080/23746149.2023.2206049},
  number    = {1},
  volume    = {8},
  journal   = {Adv. Phys. X},
  publisher = {Taylor \& Francis},
  year      = {2023},
}

@Article{barsoum2004a,
  author    = {Barsoum, M. W. and Murugaiah, A. and Kalidindi, S. R. and Zhen, T.},
  title     = {Kinking {{Nonlinear Elastic Solids}}, {{Nanoindentations}}, and {{Geology}}},
  doi       = {10.1103/PhysRevLett.92.255508},
  issn      = {0031-9007, 1079-7114},
  number    = {25},
  pages     = {255508},
  urldate   = {2026-05-06},
  volume    = {92},
  copyright = {http://link.aps.org/licenses/aps-default-license},
  journal   = {Physical Review Letters},
  langid    = {english},
  month     = jun,
  year      = {2004},
}

@Article{gottscholl2020,
  author    = {Gottscholl, Andreas and Kianinia, Mehran and Soltamov, Victor and Orlinskii, Sergei and Mamin, Georgy and Bradac, Carlo and Kasper, Christian and Krambrock, Klaus and Sperlich, Andreas and Toth, Milos and Aharonovich, Igor and Dyakonov, Vladimir},
  title     = {Initialization and Read-out of Intrinsic Spin Defects in a van Der {{Waals}} Crystal at Room Temperature},
  doi       = {10.1038/s41563-020-0619-6},
  issn      = {1476-4660},
  pages     = {540-545},
  volume    = {19},
  journal   = {Nat. Mater.},
  publisher = {Springer US},
  year      = {2020},
}

@Article{healey2023,
  author   = {Healey, A. J. and Scholten, S. C. and Yang, T. and Scott, J. A. and Abrahams, G. J. and Robertson, I. O. and Hou, X. F. and Guo, Y. F. and Rahman, S. and Lu, Y. and Kianinia, M. and Aharonovich, I. and Tetienne, J.-P.},
  title    = {Quantum Microscopy with van Der {{Waals}} Heterostructures},
  doi      = {10.1038/s41567-022-01815-5},
  issn     = {1745-2473, 1745-2481},
  number   = {1},
  pages    = {87--91},
  urldate  = {2024-07-17},
  volume   = {19},
  journal  = {Nature Physics},
  langid   = {english},
  month    = jan,
  year     = {2023},
}

@Article{onodera2020,
  author     = {Onodera, Momoko and Isayama, Miyako and Taniguchi, Takashi and Watanabe, Kenji and Masubuchi, Satoru and Moriya, Rai and Haga, Taishi and Fujimoto, Yoshitaka and Saito, Susumu and Machida, Tomoki},
  title      = {Carbon Annealed {{HPHT-hexagonal}} Boron Nitride: {{Exploring}} Defect Levels Using {{2D}} Materials Combined through van Der {{Waals}} Interface},
  doi        = {10.1016/j.carbon.2020.05.032},
  issn       = {0008-6223},
  pages      = {785--791},
  urldate    = {2026-02-17},
  volume     = {167},
  journal    = {Carbon},
  month      = oct,
  shorttitle = {Carbon Annealed {{HPHT-hexagonal}} Boron Nitride},
  year       = {2020},
}

@Article{chen2019b,
  author  = {Chen, Chunlin and Yin, Deqiang and Kato, Takeharu and Taniguchi, Takashi and Watanabe, Kenji and Ma, Xiuliang and Ye, Hengqiang and Ikuhara, Yuichi},
  title   = {Stabilizing the Metastable Superhard Material Wurtzite Boron Nitride by Three-Dimensional Networks of Planar Defects},
  doi     = {10.1073/pnas.1902820116},
  issn    = {0027-8424, 1091-6490},
  number  = {23},
  pages   = {11181--11186},
  urldate = {2026-05-26},
  volume  = {116},
  journal = {Proceedings of the National Academy of Sciences},
  month   = jun,
  year    = {2019},
}

@Article{gottscholl2021,
  author  = {Gottscholl, Andreas and Diez, Matthias and Soltamov, Victor and Kasper, Christian and Sperlich, Andreas and Kianinia, Mehran and Bradac, Carlo and Aharonovich, Igor and Dyakonov, Vladimir},
  title   = {Room Temperature Coherent Control of Spin Defects in Hexagonal Boron Nitride},
  number  = {eabf3630},
  volume  = {7},
  journal = {Sci. Adv.},
  year    = {2021},
}

\end{document}